\documentclass[useAMS,usenatbib]{mn2e}
\usepackage{graphicx}
\usepackage{txfonts} 

\newcommand{\td}{\ensuremath{t}}
\catcode`\@=11
\def\gsim{\ifmmode{\mathrel{\mathpalette\@versim>}}
    \else{$\mathrel{\mathpalette\@versim>}$}\fi}
\def\lsim{\ifmmode{\mathrel{\mathpalette\@versim<}}
    \else{$\mathrel{\mathpalette\@versim<}$}\fi}
\def\@versim#1#2{\lower 2.9truept \vbox{\baselineskip 0pt \lineskip
    0.5truept \ialign{$\m@th#1\hfil##\hfil$\crcr#2\crcr\sim\crcr}}}
\catcode`\@=12
\newcommand{\msun}{\ensuremath{\rm M_\odot}}

\def\pn{\par\noindent}
\def\mpb{\medskip\pn$\bullet$\quad}

\begin{document}
\title[Rates, Progenitors and Cosmic Mix of Type Ia Supernovae]{Rates, 
Progenitors and Cosmic Mix of Type Ia Supernovae} 
\author[Laura Greggio, Alvio Renzini and Emanuele Daddi]{Laura
Greggio$^{1}$\thanks{E-mail: laura.greggio@oapd.inaf.it (LG);
alvio.renzini@oapd.inaf.it (AR); edaddi@cea.fr (ED)}, Alvio
Renzini$^{1}$\footnotemark[1] and Emanuele Daddi$^{2}$\footnotemark[1] \\
 $^{1}$INAF - Osservatorio
Astronomico di Padova, Vicolo dell'Osservatorio 5, I-35122 Padova,
Italy\\
 $^{2}$Laboratoire AIM, CEA/DSM-CNRS-Universit\'e Paris Diderot, Irfu/SAp, 
Orme des Merisiers, F-91191, Gif-sur-Yvette, France}

\date{Accepted ... . Received 1....; in original form}
 \pagerange{\pageref{firstpage}--\pageref{lastpage}} \pubyear{2002}

\maketitle
                                                            
\label{firstpage}

\begin{abstract}
Following an episode of star formation, Type Ia supernova events occur
over an extended period of time, following a distribution of delay
times (DDT). We critically discuss some empirically-based DDT
functions that have been proposed in recent years, some favoring very
early (prompt) events, other very late (tardy) ones, and therefore
being mutually exclusive. We point out that in both cases the derived
DDT functions are affected by dubious assumptions, and therefore there
is currently no ground for claiming either a DDT strongly peaked at
early times, or at late ones. Theoretical DDT functions are known to
accommodate both prompt as well as late SNIa events, and can account
for all available observational constraints.  Recent observational
evidence exists that both single degenerate and double degenerate
precursors may be able of producing SNIa events.  We then explore on
the basis of plausible theoretical models the possible variation with
cosmic time of the mix between the events produced by the two
different channels, which in principle could lead to systematics
effects on the SNIa properties with redshift.
\end{abstract}

\begin{keywords}
supernovae: general -- galaxies: evolution -- galaxies: high redshift
\end{keywords}


\maketitle


\section{Introduction}
Type Ia supernovae (SNIa) play a prominent role in current cosmology
and astrophysics. They have been instrumental in discovering the
acceleration in the expansion of the universe (Riess et al. 1998;
Perlemutter et al. 1999), are major producers of iron in galaxies and
clusters of galaxies (e.g., Matteucci \& Greggio 1986; Renzini 1997;
B\"ohringer et al. 2004; Sato et al. 2007), contribute to power
winds/outflows in elliptical galaxies (Ciotti et al. 1991; Ciotti \&
Ostriker 2007), and their precursors may contribute to the UV and soft
X-ray emission from elliptical galaxies (Greggio \& Renzini 1990; van
den Heuvel et al. 1992).

Given the importance of SNIa's, the identification of the variety of
their precursors and the evolution of their rate following an episode of
star formation have become central issues whose relevance goes well
beyond the field of supernova physics. 

There is general agreement that SNIa events arise for the thermonuclear
explosion of a white dwarf (WD) whose mass increases until explosive
carbon burning is ignited. Two classical scenarios have
been proposed for the SNIa precursors, which differ in the mode of WD
mass increase: the so called {\it single degenerate} (SD) model of
Whelan \& Iben (1973), in which the WD accretes and burns hydrogen-rich
material from a companion, and the {\it double degenerate} (DD)
option, in which the merging of two WDs in a close binary triggers the
explosion (Webbink 1984; Iben \& Tutukov 1984).

For a long time there has been no direct observational evidence
favoring one scenario over the other, but exciting developments have
taken place recently.  A radial velocity survey of $\sim 1000$ nearby
WDs has shown that over 10\% of all WDs are indeed DD systems, and
that systems with a combined mass close to or exceeding the Chandrasekhar
limit not only exist, but can be close enough to merge in less than
one Hubble time due to the emission of gravitational waves (Napiwotzki
et al. 2002, 2003). Very recent is the discovery of variable
circumstellar absorption lines in the SNIa 2006X, which indicates the
SD nature of its precursor (Patat et al. 2007). Finally, the tentative
association of an X-Ray source near the position of (and prior to the
explosion of) SNIa 2007on (Voss \& Nelemans 2008), may represent the
first direct detection of the precursor of a SNIa event. Since WDs
that accrete and burn hydrogen are expected to be super-soft X-Ray
sources, Voss \& Nielemans favor the SD scenario.  However, the DD
scenario cannot be ruled out for this event, given that a hot (hence
X-ray emitting) accretion disk is likely to be the intermediate
configuration of the system, between first WD-WD Roche-lobe contact
and explosion (Yoon et al. 2007). Thus, quite plausible candidates for
both SD and DD scenarios exist, indicating that nature may well
feed both channels (see also Parthasarathy et al. 2007).

Besides the nature of the precursors, the second critical issue
concerns the evolution of the SNIa rate past an episode of star
formation, which is proportional to the so-called distribution of the
delay times (DDT) between the birth of the precursor and the
explosion.  The SD model predicts a sharp rise of the SNIa rate,
starting from zero at a delay time $\td \simeq 3\times 10^7$ yr,
reaching a maximum $\sim 10^8$ yr later, then followed by a steady
decline extending for about one Hubble time or more (Greggio \&
Renzini 1983). A quite similar behavior is found in the so-called {\it
scenario code} rendition of the DD channel (Yungelson et al. 1994),
whereas Greggio (2005) has demonstrated that for a very wide range of
adopted parameters a sharp rise to the maximum, followed by a (quasi)
plateau lasting a few $10^8$ to a few $10^9$ yr, and by a steady,
long-lasting decline is a generic property of both the SD and the DD
models. This is illustrated in Fig 1 showing a selection of model
DDTs from Greggio (2005) for both the SD and the DD cases. These
distributions depend on various parameters such as the mass range of
the components of the binary systems which result into a successful
explosion, or the amount of orbital shrinkage during binary
common envelope evolution, which affects the distribution of the
separations of the DD systems at birth.  For the SD model the late
epoch steep decline is related to the need of securing a Chandrasekhar
mass by summing the WD mass and the envelope of a progressively lower
mass donor, which constrains the initial mass of the primary in a
progressively narrower range. For the DD models the explosion time is set by
the evolutionary timescale of the secondary plus the time taken by
the DD system to merge by gravitational wave radiation, which
is a strong function of the DD separation. The 
options DD WIDE and DD CLOSE refer to two different assumptions for the degree
of orbital shrinkage during the first common envelope phase, with the DD CLOSE
case favoring events with short delay times.
For more details see Greggio (2005).

At present existing data do not allow a firm preference for one model
over another, but we note that the behavior of their DDTs quite
naturally predicts both a high rate in blue, star-forming galaxies, as
well as some level of SNIa activity in red, passively evolving
ellipticals (Greggio 2005), which is in fine agreement with state of
the art estimates of the SNIa rate in local galaxies of various
morphological types (Cappellaro, Evans \& Turatto 1999; Mannucci, et
al. 2005).

\begin{figure}
\includegraphics[width=84mm]{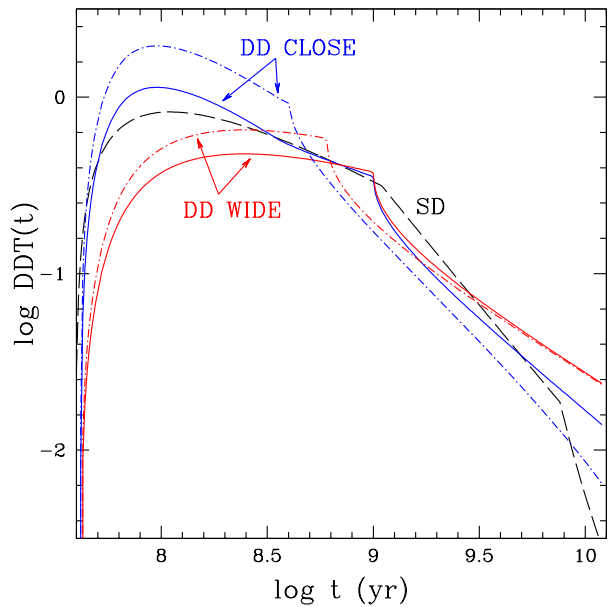} 
\caption{Examples of the distribution of SNIa delay times for the 
SD and the DD models, adapted from Greggio (2005). The SD model 
(long dashed line) 
assumes a flat distribution of the mass ratios of the binary progenitors, 
and a full efficiency of WD mass growth relative to the mass transferred by the
donor. Two realizations of the DD CLOSE, and DD WIDE models are shown,
all relative to a flat distribution of the separations of the DD 
systems. These models assume different values for the 
minimum mass of the secondary in the binary SNIa
progenitor: 3 and 2 \msun \ respectively for the dot dashed and the solid 
DD CLOSE cases; 2.5 \msun \ and 2 \msun respectively for the dot dashed and the
solid DD WIDE lines. 
The DDTs are normalized to 1 over the range $0.03 \leq \td \leq
12$ Gyr, so to give the same total number of events.}
\label{DDTs}
\end{figure}

For a variety of astrophysical applications, conveniently simple
parameterizations of the DDT function have been often adopted, such as
e.g., a declining power law ($DDT\propto \td^{-\alpha}$, Ciotti et
al. 1991; Renzini 1997), or an exponential ($DDT\propto
e^{-\td/\tau}$, Madau, Della Valle \& Panagia 1998).  Given the
growing interest on SNIa's, more empirically motivated renditions of
the DDT/SNIa rate have been recently proposed which are clearly
irreconciliable with each other. For example, the DDT function
proposed by Strolger et al. (2004) implies no SNIa events at all
during the first $\sim 2$ Gyr, whereas $\sim 50\%$ of all events would
take place during the first $\sim 10^8$ yr according to the DDT
function favored by Mannucci, Della Valle \& Panagia (2006). These
latter authors also appeal to a {\it bimodal} SNIa rate/DDT, with a
{\it prompt} component (possibly) separated from a {\it tardy} ones, a
scenario that is becoming quite popular.

In this paper we examine critically these proposed SNIa distributions
of delay times, first checking their empirical basis, then discussing
whether they can be reconciled (or not) with their theoretical
counterparts, and whether they are consistent with other relevant
evidences in a broader astrophysical context. Finally, we discuss to
which extent these recent observational developments lead to progress
in our understanding of the nature of SNIa precursors and the
evolution of their rate through cosmic times. The current concordance
cosmology is adopted, when needed ($H_\circ =70,\; \Omega_{\rm
M}=0.3,\;\Omega_\Lambda=0.7$).

\section{DDT function motivated by the redshift evolution of the SNIa rate}
Strolger et al. (2004, 2005) have proposed for the DDT function a
Gaussian peaking at $\td=3.4$ Gyr after the birth of the star
progenitors, with $\sigma = 0.68$ Gyr, which has virtually no SNIa
events during the first $\sim 2$ Gyr, i.e., a scenario lacking
completely the {\it prompt} events. This result is at macroscopic
variance with theoretically-motivated DDT functions, as well as with
other empirically motivated functions to be discussed in the next
sections.  In particular, such a Gaussian function cannot reproduce
the trend of SNIa rate with galaxy type/colour, which is highest in
actively star-forming galaxies (Greggio 2005; Mannucci et al. 2006;
Sullivan et al. 2006).  For these reasons, we re-discuss here this
result and its empirical basis.

Strolger et al. DDT function is based on the SNIa rates as function of
redshift derived by Dahlen et al. (2004) from the supernova survey
spin-off of the GOODS project (Giavalisco et al.  2004).  The
mentioned Gaussian DDT is a consequence of the high peak of the SNIa
rate at $z \sim 0.8$, followed by its sharp drop at $z>1.4$.  The
height of the peak, first confirmed by Barris and Tonry (2006), has
been later questioned by other measurements. Actually, in the current
literature there is a large discrepancy between the estimates of the
SNIa rate at $z$ between 0.5 and 0.8 (see recent compilation in Blanc
and Greggio 2008), with other determinations lying formally more than
$\sim 2\sigma$ below the SNIa rates of Barris and Tonry (2006) and
Dahlen et al. (2004).  If the latter measurements are excluded, the
observed trend of the SNIa rate with redshift can be well accounted
for with wide DDTs encompassing both prompt and tardy events
such as those shown in Fig. 1 (Botticella et al. 2008, Blanc and
Greggio 2008).

However, it is the low rate at $z=1.6$ measured by Dahlen et al. (see
also Poznanski et al. 2007 and Kuznetsova et al. 2008) that basically
requires setting to zero the DDT function for $\td\lsim 2$ Gyr, hence
dropping completely the {\it prompt} events.  In Dahlen et al. the
data point at $z = 1.6$ is derived from just 2 events (3 events in
Poznanski et al. 2007, but with more uncertain redshifts), indeed a
regime of small number statistics. However, somewhat more
statistically significant are the {\it missing} events that would be
expected from DDT functions that peak at $\td \sim 10^8$ yr, such as
those e.g., shown in Fig. 1. Scaled to the Dahlen et al. sample, these
DDT functions would predict $\sim 4-6$ events in the $z=1.6$ redshift
bin, whereas only 2 are observed.

One critical assumption in Dahlen et al. concerns the extinction, for
which $E(B-V)=0.15$ is adopted for all galaxies and all
redshifts. However, as pointed out by e.g., Mannucci, Della Valle
\& Panagia (2007), star forming galaxies at high redshift are likely to
be affected by higher extinction than local ones, 
an issue we quantify below. Indeed,
extinction must become more and more important at higher and higher
redshifts for several reasons:

\mpb Supernovae are discovered on one-band images, hence sampling
shorter rest-frame wavelengths the higher the redshift. Thus, the
higher the redshift, the stronger the extinction in the rest frame
band, since extinction increases with decreasing wavelength, such as
in the extinction law of Calzetti et al. (2000) for which $A_{\rm V} =
4\times E(B-V)$, and $A_{1500}=10\times E(B-V)$.  In the case of the
HST/GOODS survey, supernovae are discovered on $z$-band (F850LP)
images, hence sampling the rest frame $\sim 5700$ \AA\ at $z=0.5$,
$\sim 4200$ \AA\ at $z=1$ and $\sim 3200$ \AA\ at $z=1.6$.

\mpb In a given SN survey, the extinction correction has a relatively
small effect at low redshift, since events are quite bright and
discovered well above the magnitude limit for detection. Instead, the
effect of extinction increases with redshift as events approach the
detection limit, and becomes dominant near the limit redshift reached
by the survey. In this way, an underestimate of the extinction
(assumed independent of redshift) will have only a modest  effect on the
estimated SN rate at low redshift, but will have a very large effect
near the redshift limit of the survey, both because the
extinction correction is largest at the shortest wavelength sampled in
the rest frame, and because it will cause these dimmest supernovae to fall
below the detection limit.

\mpb These two effects are further magnified if the average extinction
(at fixed rest frame wavelength) increases with redshift, with the
result of systematically biasing the estimated SN rate, recovering
only the least obscured events. Again, the effect is maximum in the
highest redshift bin.

\begin{figure}
\includegraphics[width=84mm]{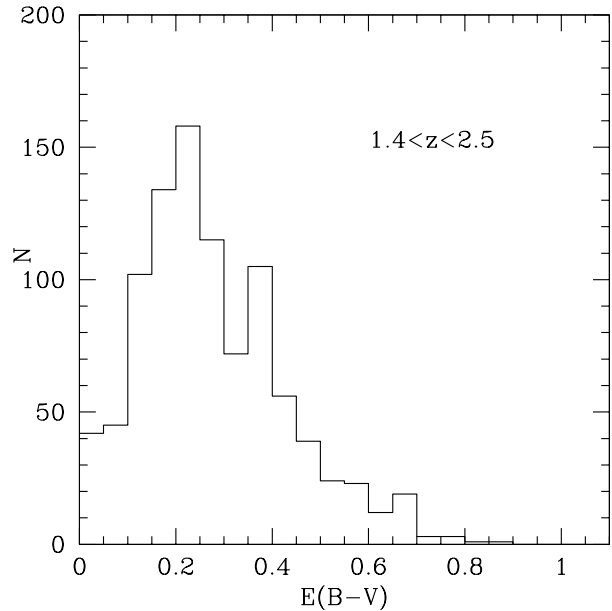}
\caption{The distribution of reddening among $K_{\rm Vega}<22$ galaxies at 
$1.4<z<2.5$ in the GOODS-S field (from Daddi et al. 2007).
.}
\label{isto}
\end{figure}

\medskip

A direct estimate of the distribution of reddening/extinction, based
on the observed UV slopes, was obtained by Daddi et al. (2007) for the
$\sim 1000$ star-forming, $K_{\rm Vega}<22$ galaxies at $1.4<z<2.5$ in
the GOODS-South field, with $E(B-V)$ set to zero for passively
evolving galaxies.  As generally adopted in the study of high redshift
galaxies, the extinction law by Calzetti et al. (2000) was used. The
$K$-band limit of this sample corresponds to a $M\gsim
4\times10^9M_\odot$ stellar mass limit.  To this level, we account for
a stellar mass density of $5.3\times10^{7}M_\odot$~Mpc$^{-3}$, about
2/3 of the total at $z=1.8$, integrated over a Schechter function fit
in the GOODS-South sample (Fontana et al. 2006). This sample also
accounts for a star-formation rate density of
0.12~$M_\odot$~yr$^{-1}$~Mpc$^{-3}$, close to the (admittedly
uncertain) integrated value at $z\sim1.5$--2 (e.g., Perez-Gonzalez et
al. 2005; Hopkins \& Beacom 2006).  A straight Salpeter IMF is used
for these derivations and comparisons. Thus, we believe that this
sample of galaxies includes the bulk of the stellar mass and star
formation at these redshifts, hence the bulk of SNIa events.
Note also that the reddening $E(B-V)$ derived from the UV slope is, in
case, biased towards the less obscured regions of a starforming galaxy,
since those which have a higher extinction give a small (if any) 
contribution to the UV flux, hence to the UV slope.

\begin{figure}
\includegraphics[width=84mm]{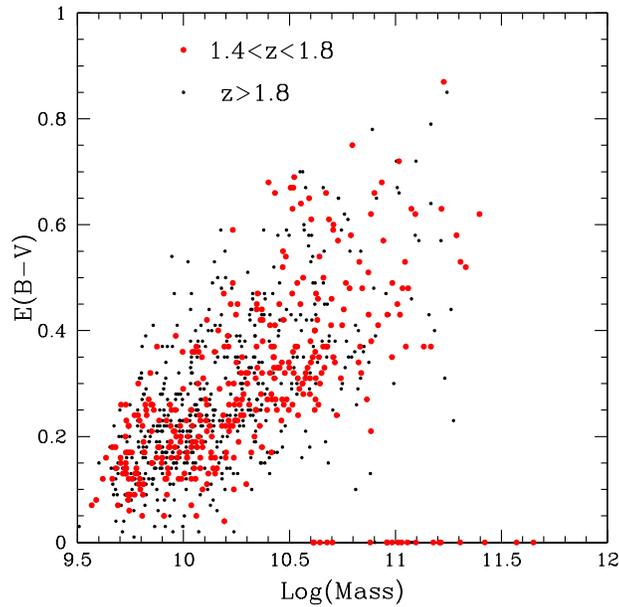}
\caption{The reddening vs stellar mass (in $\msun$ units) for the
$K_{\rm Vega}<22$ galaxies at $1.4<z<2.5$ in the GOODS-S field (from
Daddi et al. 2007). Galaxies at $1.4<z<1.8$ and $z>1.8$ are plotted
with different symbols as indicated in the figure.  Note the small
group of massive, passively evolving galaxies with $E(B-V)=0$.  }
\label{MASS}
\end{figure}

\begin{figure}
\includegraphics[width=84mm]{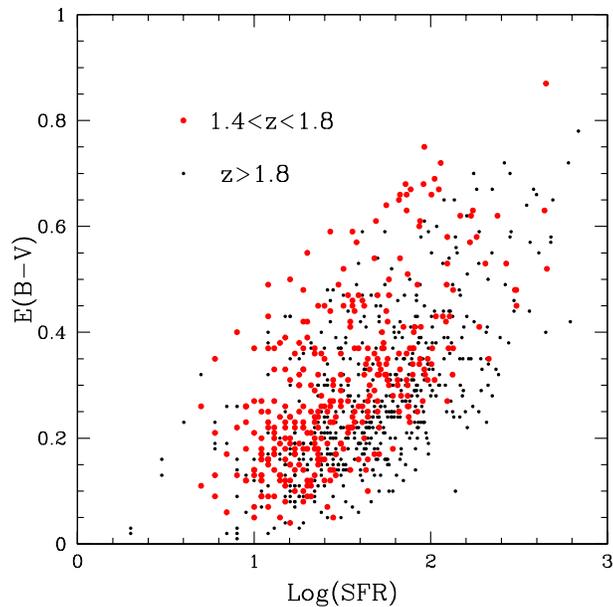}
\caption{The reddening vs. star formation rate (in $\msun\;{\rm
yr}^{-1}$) for the $K_{\rm Vega}<22$ galaxies at $1.4<z<2.5$ in the
GOODS-S field (from Daddi et al. 2007).}
\label{SFR}
\end{figure}

\begin{figure}
\includegraphics[width=84mm]{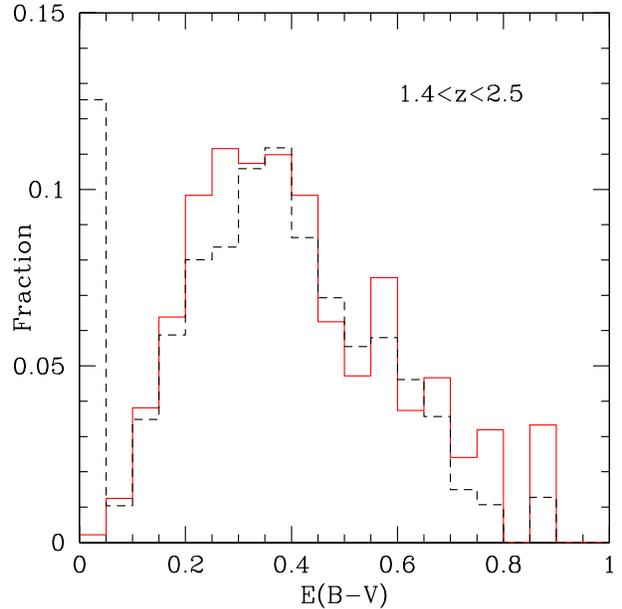}
\caption{The fraction of stellar mass (dashed histogram) and of the
star formation rate (full line histogram) in each reddening bin for the same
sample of $1.4<z<2.5$ galaxies shown in Fig. 2, 3 and 4. Note the high
spike at $E(B-V)=0$ in the mass distribution, which is due to the
passively evolving galaxies. Correspondingly, no such spike is present
in the star formation rate histogram.}
\label{SFR1}
\end{figure}

Fig. 2 shows the resulting reddening distribution, which peaks at
$E(B-V)=0.2$, not much larger than the average value adopted by Dahlen
et al. However, what matters for the SN statistics are primarily
the mass-weighted and SFR-weighted average values of $E(B-V)$. Fig. 3
and 4 show, respectively, the reddening $E(B-V)$ vs. the stellar mass
and the SFR for each of the galaxies in the GOODS-S sample.  The
reddening is definitely stronger in more massive, more star-forming
galaxies, i.e., in those galaxies that are more productive of SNIa
events. Indeed, the mass-weighted and SFR-weighted reddening $E(B-V)$
are respectively 0.36 and 0.41. Fig. 5 shows both the fraction of the
stellar mass and of the star formation rate in each reddening bin.
Note that $\sim 12.5\%$ of the stellar mass at these redshifts is in
passively evolving galaxies with $E(B-V)=0$. Thus, we adopt
$E(B-V)\sim 0.4$ for the average reddening affecting SNIa events at
these redshifts, as to a first approximation the SNIa rate depends on
a simple linear combitation of galaxy stellar mass and current star
formation rate (see Section 3).

As mentioned above, at $z=1.6$ the F850LP filter samples the 3200 \AA\
in the rest frame, and for the Calzetti et al. (2000) extinction law
one has $A_{3200}\simeq 7\times E(B-V)$, i.e., an average $\sim 2.8$
magnitudes of extinction for $<\!  E(B-V)\!>=0.4$.  Adopting instead
$E(B-V)=0.15$ at all redshifts, as in Dahlen et al. (2004), one
derives a much smaller average extinction, i.e., $\sim 1$ magnitude at
$z=1.6$. Therefore, we argue that Dahlen et al. have underestimated by
almost two magnitudes the mass/SFR averaged extinction of galaxies at
$z=1.6$, and have correspondingly underestimated the SNIa rate by a
large factor.  A quantitative estimate of this effect can be
attempted by following the Dahlen et al. prescription according to
which a 0.06 error in $E(B-V)$ generates a 1$\sigma_{\rm stat}$ error
in the estimated SNIa rate. Thus, increasing $<\!  E(B-V)\!>$ from
0.15 to 0.40, the derived SNIa rate at $z=1.6$ is going to be higher
by $\sim 4\sigma_{\rm stat}$.  With the statistical error $\sigma_{\rm
stat}$ from Table 2 in Dahlen et al. this corresponds to a factor
$\sim 4$, and therefore by nearly as much was the SNIa rate
underestimated. Of course, this is quite a rough estimate of the
effect, as the derived rate of SN events does not depend only on the
average extinction, but specifically on the full distribution of
extinctions (such as shown in Fig. 5), and even on the distribution of
extinction within each individual galaxy relative to the observer.

More recently, Dahlen, Strolger \& Riess (2008) have analysed an
expanded sample of SNIa events over the GOODS fields, bringing the
total to 56 events compared to the 23 events in the original GOODS
sample. However, the number of events in the highest redshift bin
$z=1.4-1.8$ increases only from 2 to 3. All in all, Dahlen et
al. (2008) confirm their previous result, i.e., a low rate at $z=1.6$,
having adopted nearly the same assumptions concerning the average
extinction as in Dahlen et al. (2004), independent of galaxy mass,
star formation rate and redshift. In their recent study, Dahlen et
al. discuss the effect of various extinction distributions relative
merely to inclination effects, including some with a tail to very high
extinctions. All such assumed distributions peak at zero extinction,
at variance with the pertinent empirical distributions shown in Fig. 2
and 5. Of course, what matters is the distribution of extinctions
affecting specifically the SNIa events, which may differ from the
distribution of average extinctions affecting galaxies as a
whole. However, we conclude that Dahlen et al. (2004, 2008), having
strongly underestimated the average extinction of high redshift
galaxies, are likely to have underestimated by a large
factor the SNIa rate at $z=1.6$.  Therefore, there remains little
ground for the extremely {\it tardy} DDT function inferred by Strolger
et al. (2004, 2005).

\section{The two-component DDT}
Scannapieco \& Bildsten (2005) propose a crude, yet effective
representation of the SNIa rate in galaxies, with a component
proportional to the mass in stars, and another to the current SFR:

\begin{equation}
R_{\rm Ia}(t) = AM_*(t) + B\dot M_*(t).
\end{equation}

In practice, the implied DDT(\td) is the sum of a Dirac's delta at
$\td=0$ and a constant, representing respectively a {\it prompt} and a
{\it tardy} component, i.e., $DDT(\td)=\delta(0)+{\rm
const}$. Counting on the two adjustable parameters $A$ and $B$, this
simple rendition of the DDT function can quite satisfactorily
reproduce several SN related properties of stellar systems, such as
e.g., the correlation of the SNIa rate with galaxy colour in the local
universe, the SNIa rate history of star-forming galaxies, the origin
of the $\alpha$-element enhancement in spheroids, etc.. However, it
certainly fails in other astrophysical situations. For example, it
implies a non-evolving SNIa rate in passively evolving galaxies, in
which the cumulative number of SNIa's increases linearly with time,
hence a diverging productivity of SNIa's for a finite mass in
stars. At the same time, passively evolving (SFR=0)
galaxies of given mass, but vastly different ages would have the same
SNIa rate according to Eq. (1), contrary to the empirical evidence in
Totani et al. (2008). Therefore, formulation (1) can only be applied 
to a limited number of concrete situations, and can produce 
invalid results in others.

The SNIa rate in galaxies is the convolution of the DDT with
the star formation history (SFH):

\begin{equation}
R_{\rm Ia}(t) = \int_{0}^{t} {\rm SFR}(t-t')\, {\rm DDT}(t')\, {\rm d} t' 
\end{equation}

\noindent
having indicated with $t'$ the delay time. By splitting the
integration over short ($prompt$ component) and long delay times
($tardy$ component), this equation reduces to Eq. (1), with B
proportional to the fraction of $prompt$ events, and A proportional to
the SFR-weighted-DDT, averaged over the $tardy$ events. Hence, when
fitting relation (1) to a set of data, $B$ will depend on the time
span assumed for the prompt events, and $A$ on the adopted law for the
SFH (besides the shape of the DDT).  Indeed, very different values for
the $A$ and $B$ {\it constants} are found in the literature \textbf
(Scannapieco \& Bildsten 2005; Sullivan et al. 2006; Neill et
al. 2006), with with $B$ values that differ up to a factor of $\sim
5$.  Whereas part of the discrepancies are due to the different
quality of the various data samples, the application of Eq. (1)
naturally results into a dependence of $A$ and $B$ on the (arbitrary)
choice of the time delay separating $prompt$ from $tardy$ events, and
on the actual SFH of the galaxies in the samples. This is to say
that $A$ and $B$ cannot be universal constants.
%


The shortcoming with relation (1) is
that it does not adequately describe the contribution to SNIa events
at intermediate delay times, and the prompt events are completely
attributed to the ongoing SF. While this is correct in the case of
core collapse (CC) SNe, there's no reason to hold this true for SNIa
as well.  The existence of a channel for SNIa closely related to the
current SFR was also suggested in Mannucci et al. (2005), having
noticed that in star forming galaxies the SNIa rate varies nearly in
locksteps with the rate of CC supernoave.  Actually, the ratio of the
CC to the SNIa rates increases as the galaxies get bluer or of later
type, especially from Sa to Sc (Mannucci et al. 2005), indicating that
also in very actively star-forming galaxies SNIa events lag behind CC
events.

\section{DDT motivated by the SNIa rate in radiogalaxies}

Mannucci et al. (2006) have advocated a {\it bimodal} DDT, with the
{\it prompt} and {\it tardy} components being physically distinct or
even separated in time. The driving argument for the bimodality
follows from the excess SNIa events that is observed in a sample of
radio-loud elliptical galaxies, with respect to radio-quiet galaxies
of the same morphological class (Della Valle et al. 2006). This is a
$\sim 2\sigma$ excess, that Della Valle et al. ascribe to star formation
having been triggered by the same merging/accretion event powering the
nuclear activity in radio-loud galaxies, whose duration is assumed to be 
$\sim 10^8$ years. The consequence of this
interpretation, is a favored DDT function in which $\sim 50\%$ of all
SNIa's explode within the first $\sim 10^8$ yr after a star formation
episode (the prompt component), and the rest are distributed over the
following many Gyr (the tardy component). Taken quantitatively at face
value, this inferred DDT function would have quite important
consequences for constraining the nature of the progenitors, excluding
most of otherwise plausible models, see for example Fig. 13 in Greggio
(2005).  For this reason, we take the liberty of re-discussing here
the Della Valle et al. interpretation of the SNIa excess in radio-loud
ellipticals.

We first quantify this excess. In the sample analysed by Della Valle
et al., 7 SNIa's were found among the radio-quiet ellipticals, and 9
among the radio-loud ones, with an additional SN having exploded in a
galaxy with borderline radio power, hence it was attributed for one
half to the radio-quiet group and one half to the radio-loud group.
The relative normalized control times (the product of the time
during which a galaxy was monitored times its $B$-band luminosity)
were cumulatively 7127 and 2199 (yr$\times L^{\rm
B}_\odot/10^{10}L^{\rm B}_\odot$), respectively for the sample of
radio-quiet and radio-loud galaxies. Thus, if the productivity of SNIa
events was the same regardless of radio power, one would have expected
to find $\sim 2199\times 7.5/7127=2.3$ SNIa's in the radio-loud
sample, while instead 9.5 have been observed. There was therefore a
sizable excess of $\sim 7$ events, that Della Valle et al. attribute
to a recent star formation episode. In terms of SN rates, they infer a
SNIa rate of $0.43^{+0.19}_{-0.14}$ SNU for the radio-loud galaxies,
and $0.11^{+0.06}_{-0.03}$ SNU for the radio-quiet ones (1 SNU = 1 SN
per century per $10^{10}L^{\rm B}_\odot$).

Della Valle et al. (2006) argue that if the excess SNIa events is due
to recent star formation in radio-loud galaxies, then they should be
accompanied by a corresponding share of core collapse (CC) supernovae,
given that in star forming galaxies CC supernovae outnumber the SNIa's
8 to 3. However, no CC supernova has been found in the whole sample of
radio-loud galaxies. Della Valle et al. argue that this is not in
contradiction with their star formation hypothesis, since they
calculate that only less than one (i.e., 0.62) CC supernova should
have been found. We believe that their calculation is not correct, and
repeat it here, based on the same input numbers. We estimate the
number of CC supernovae that should have resulted from the same star
formation event responsible for the excess SNIa's in the following
way:
\begin{equation}
N_{\rm CC}= {R_{\rm CC}\over R_{\rm Ia}}\cdot N_{\rm Ia}\cdot {CT_{\rm CC}\over
            CT_{\rm Ia}},
\end{equation}
where the ratio of the two SN rates is taken as 8/3, the number $
N_{\rm Ia}$ of excess SNIa's is 7, and the ratio of the two control
times is 4057/2199=1.84 in favor of CC supernovae. The result is that
34 (!) CC supernovae should have exploded, but none was observed.  The
discrepancy with the $\sim 30$ times smaller estimate by Della Valle
et al. stems from them having multiplied this number (34), by 0.03,
the ratio of the maximum lifetime of CC supernova progenitors (30 Myr)
to an assumed duty cycle between successive episodes of nuclear radio
activity (1 Gyr). Multiplication by this factor is not appropriate,
as neither the lifetime of CC supernova progenitors, nor the
radiogalaxy duty cycle have any connection with the observability of
the predicted 34 events.  Moreover, the 8/3 CC/SNIa ratio pertains to
steadily starforming galaxies, in which the full range of delay times
contribute to the SN rates. In this particular case, the 7 {\it
excess} SNIa events are assumed to belong to an extremely prompt
component, that would contribute some $\sim 50\%$ of all events, and
therefore the appropriate CC/SNIa ratio should actually be twice as
large, i.e., $\sim 16/3$, implying that the 7 SNIa events should have
been accompanied by $\sim 69\pm 26$ CC events. The Della Valle et
al. sample of 267 radio loud galaxies can be regarded as a unique
entity having experienced continous star formation over the last $\sim
10^8$ years, even if star formation occurred in bursts within
individual galaxies. Thus, Eq. (3) (without further reducing factors)
gives the total number of CC events that should have been observed
within the sample of radiogalaxies.  Failing to detect so many CC
events could result if galaxies in the act of a starburst, or during
the on phase of CC supernovae, would not qualify as radiogalaxies
according to the criterion adopted by Della Valle et al.  (2005), as
if radio activity would develop only after star formation (and CC
supernova events) had subsided. Instead, starburst and radio activity
are often simultaneus phenomena, as e.g., in the prototypical case of
NGC 1275 (Gallagher 2007). It seems quite unlikely to us that such decoupling
of radio activity and CC supernovae could be so extreme as to have all
predicted CC events occurring outside of the radio loud phase of galaxies.

We conclude that if the 7 excess SNIa events in the monitored
sample of radiogalaxies were due to recent star formation then a much
larger number of CC supernovae should also have been detected in the
sample of radio-loud galaxies. Since none was observed, we argue that
it is quite unlikely that the 7 events were due to starbursts that
would have occurred in the last $\sim 10^8$ years. Starbursts may well
have occurred in the sample of radiogalaxies, but likely involved
a stellar mass insufficient to produce the excess SNIa events, as also
indicated by the colours of radio-loud galaxies (in given mass bins)
being indistinguishable from those of radio-quiet galaxies (Mannucci
2007; see also Johnston et al. 2008). Recent bursts of star formation
in radiogalaxies as reponsible for their excess SNIa events is also
excluded by Cappellaro, Botticella \& Greggio (2007), based on the relative
frequency of CC and SNIa events.

Having excluded star formation, the question remains of the origin of
the observed excess. One quite plausible possibility is just small
number statistics: after all the excess is only a 2-$\sigma$ effect,
and could well disappear in future, wider SN surveys.

\section{SNIa bimodality and co-existence of SD and DD progenitors}

While excluding the kind of bimodality motivated by the apparent SNIa
excess in radio-loud galaxies, recent observations suggest that a
substantially different kind of bimodality may actually exist. Indeed,
as mentioned in Section 1, there is now direct evidence for nature
using both the SD and the DD channels for producing SNIa events, although
the relative contribution of the two channels remains substantially
unconstrained. The DDT functions for the SD and DD cases depend on
partly different physical ingredients: the nuclear lifetime of the
secondary star in the binary progenitor for the SD case, with the
addition of the gravitational wave radiation delay for the DD case.
Therefore, following a star formation episode, the relative contributions
of the two channels is going to change with time, and so will their
mix as a function of cosmic time when convolved with the cosmic
history of star formation.

\subsection{The evolution of the SNIa mix}

In this section we explore how the relative importance of DD and SD
events may vary with the age of a stellar population, as well as over
the cosmic time. If the two SNIa channels produce events with
different observational characteristics (e.g., luminosity at maximum,
colour, rate of decline, etc.)  a secular variation of the ratio of
their rates impacts on the properties of observed samples, and may
potentially introduce a systematic bias on the estimated distances.
The exploration is made by using the DDT functions for the 
sample of SD and DD models shown in Fig. 1. We emphasize that our
conclusions are not affected by the specific selection of these
models, but rather reflect their generic properties.

\begin{figure}
\includegraphics[width=84mm]{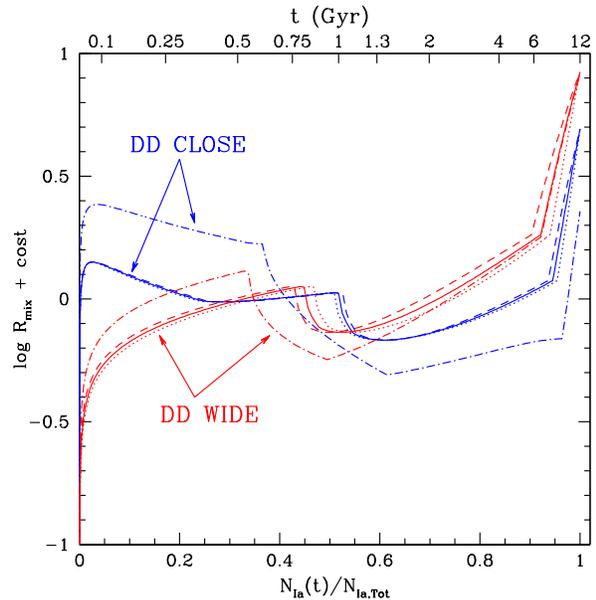}
\caption{Ratio of the rates from the DD and the SD channel as a
function of the fraction of SNIa events occurred in an SSP, for
the same models shown in Fig. 1 and with the same line encoding. The
solid and dot-dashed curves illustrate the case in which the SD and DD
channels contribute equally to the total number of SNIa events up to
$t=12$ Gyr (this corresponds to $const=0$ on the vertical axis of the
figure). The effect of a different partition between SD and DD events
is shown for the solid lines models, as dotted and dashed curves,
which plot respectively the results for a 30\% ($const$=+0.37) and
70\% ($const$=-0.37) contribution of the DD channel. Note that a different
partition essentially shifts the lines along the vertical axis, while
the effect on the horizontal axis is small. The upper axis is
labelled with the age of the SSP for the case of a 50\% contribution
from the two channels, and the DD CLOSE model shown as a solid
line. The correspondence between the age of the SSP and the fraction
of events is somewhat different for the other cases.} 
\label{RAT_SSP}
\end{figure}
\begin{figure}
\includegraphics[width=84mm]{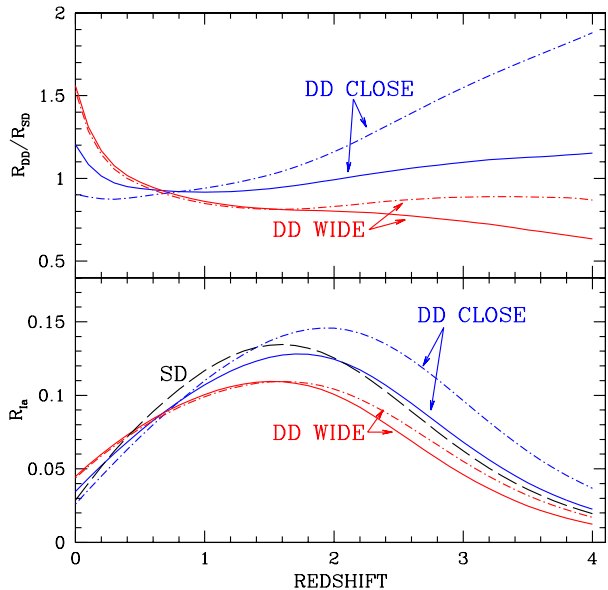}
\caption{The cosmic evolution of the SNIa rates, and of their ratios,
predicted for the model DDTs shown in Fig. 1.
The adopted cosmic star formation history is from Hopkins
and Beacom (2006), case Sal-A, Cole function up to a redshift of 6, when
star formation is assumed to start. The linetype encoding for the DD
models is the same as in Fig. 6; the bottom panel shows the evolution
for the SD model as a dashed line.}
\label{RAT_z}
\end{figure}

Fig. \ref{RAT_SSP} shows the ratio $R_{\rm mix}$ between the SNIa
rates for DD and SD models in a simple stellar population (SSP) as it
ages up to 12 Gyr. The ratio is plotted as a function of time since
the star formation episode (upper scale), as well as against the running
number of SNIa events occurred in the system up to a given time $t$
(lower scale), normalized to the total number of events up to $t=12$
Gyr.  Common to all the explored cases is that at first the SD channel
prevails, but the rate from DD explosions rapidly increases and
catches up relative to the SD events. This occurs over a time scale
which is very short ($\lsim 0.1$ Gyr) in the DD CLOSE case,
characterized by strong orbital shrinkage during the common envelope
episodes of the progenitor binary (Greggio 2005). This time is a
little longer ($\sim 0.5$ Gyr) in the DD WIDE case, where less orbital
shrinkage is assumed, and therefore gravitational wave radiation takes
longer to bring the two WDs into contact.

At later times the ratio of the two rates levels off, and resumes to
grow rapidly when the SSP ages through the second cusp in the SD DDT,
i.e. for $t\gsim 8$ Gyr (see Fig. \ref{DDTs}). This cusp is due to the
requirement of reaching the Chandrasekhar limit, which at a late epoch
remains possible only for binaries with more and more massive WD
primaries.

All in all, the ratio varies by about 2 orders of magnitude in the 0--12 Gyr
time interval, with SD events dominating at (very) early times, and DD events
largely dominating at (very) late times. However,  during most of the time
the ratio of the two rates is actually confined within $\sim\pm 40\%$
of its average value. This constant mix of the SNIa events from the two 
channels is even more so if we consider the central 90\% of the events, i.e.
cutting out the 5\% events at early times and  the 5\% events at late times.

Thus, we do not expect dramatic differences in the SD/DD mix of SNIa
samples in star-forming galaxies. However, in galaxies which have
evolved passively for a time comparable to the Hubble time, such as
elliptical galaxies, the SD-SNIa rate may have already entered its
very fast final drop seen in Fig. 1 and, correspondingly, most (if not
all) SNIa's in these galaxies may result from the DD channel which is
able to remain active for a much longer time. This would be
even more so if the accretion efficiency for the SD systems is smaller
than unity. Note that for
galaxies that have been passive over the last 10-12 Gyr, the evolution
of the SNIa rate follows closely the DDT functions for a single burst
which are shown in Fig. 1, i.e., much different from the constant rate
over time implied by the parameterization of Scannapieco \& Bildsten
(2005).

The five SNIa distributions of delay times shown in Fig. 1 have also
been convolved with the cosmic history of star formation in its
analytic rendition proposed by Hopkins and Beacom (2006). The results
are shown in Fig. \ref{RAT_z} (lower panel), along with the
corresponding evolution with redshift of the ratio between the DD and
SD rates (upper panel), in the case of an equal contribution of the
two channels to the cumulative number of events all the way to
$z=0$. If the cumulative contributions are different, the scale
changes by a multiplicative factor, but the relative evolution of the
ratio remains the same.

The behaviour of the ratio with redshift is different for the DD CLOSE
and the DD WIDE models: in the latter case, the increased importance
of the long delay time events results in a high contribution of DD
explosions at low redshift. In the DD CLOSE models, instead, the DDT
function is comparatively steeper at late times (cf. Fig. 1), so that
their contribution is less pronounced at low $z$. If the DDT is
sufficiently steep (dot-dashed line), the DD events happen to dominate
even at high redshift.

In conclusion, Fig. 7 (upper panel) shows that the expected evolution
with redshift of the SD/DD mix is rather mild. Within the range currently
explored  by observations ($z<1.8$) the largest variation is exhibited by the
combination DD-WIDE/SD, where the corresponding mix drops by $\sim 60\%$ 
between $z=0$ and 1. 

In principle, the redshift evolution of the SNIa mix may have
important implications for the systematic effects it could induce on
the distances inferred from the SNIa luminosity, and implied rate of
cosmic acceleration. Thus, the relatively modest variation with
redshift of the SD/DD mix is quite reassuring. However, we caution
that even within each of the two channels the distribution of the
properties of the SNIa progenitors are expected to change with
redshift in a systematic fashion. This is the case because at high
redshift only relatively ``prompt'' events can take place. Thus,
progenitor properties such as the age of the primary WD, or the
combined mass of the two WDs in DD systems, etc. are characterized by
distribution functions that vary with redshift, thus potentially
having an effect on the SN explosion, light curve, luminosity,
etc. For example, the age of the primary WD controls its internal
temperature stratification, and the extent to which crystallization
and/or carbon-oxygen diffusive separation have taken place inside the
WD prior to the SN explosion. In turn, these properties of the
progenitor system may affect the explosion itself, hence the
light-curve, and indeed it is well known that SNIa's exhibit a range
of peak luminosities.  It is also well known that peak luminosity and
shape of the light curve (specifically, it rate of decline) correlate
closely, and therefore these variations are corrected for when using
SNIa's for distance determinations (e.g., Riess 1998). However, it
remains to be empirically established whether the local calibration of
the relation between luminosity and rate of decline holds true also at
high redshifts.

\subsection{Predicted evolution with redshift of the SNIa rate}

Fig. 7 (lower panel) shows the evolution with redshift of the SNIa
rate for each of the 5 DDT functions shown in Fig. 1, having convolved
the latter ones with the cosmic SFH from Hopkins \& Beacom (2006).
The generic expectation from both SD and DD models is an increase of
the rate with redshift by a factor 3--4 between $z=0$ and $z\sim 1$,
in fair agreement with current empirical estimates, e.g. as reported
by Dahlen et al. (2008). However, this increasing trend continues up
to $z=1.6-2$, while the empirical rate by Dahlen et al. (2004, 2008)
drops to a low level at $z=1.6$. Note that a peak SNIa activity at
$z\sim 2$ is predicted also by other SD models for the SNIa
precursors, that include as possible precursors WD+ main sequence star
systems (Kobayashi, Tsujimoto \& Nomoto 2000).  The reasons for the
apparent drop at $z=1.6$ have been already discussed, and we believe
that the key issue is the extinction in real galaxies at this redshift
being $\sim 2$ magnitudes higher in the rest frame than assumed by
Dahlen et al.. Deeper $z$-band observations would be needed to
vindicate or exclude the SD/DD model predictions for $z>1.4$ shown in
Fig. 7. Even better, deep near-IR observations would sample rest-frame
wavelengths less affected by internal extinction. A first opportunity
will still be offered by HST, which after the next Servicing Mission will
provide deep $H$-band imaging with WFC3, thus probing the rest-frame
$\sim 6000$ \AA \ at $z=1.6$, and $\sim 5300$ \AA \ at $z=2$.
Exploring the declining SNIa rate beyond redshift $\sim 2$ should be
possible with future JWST observations at wavelengths longer than
$2\;\mu$m, that at these redshifts will sample the relatively
unobscured rest-frame $I$ band.

\section{Conclusions}

We have examined the empirical basis for the very {\it tardy}
distribution of SNIa delay times advocated by Strolger et al. (2004),
derived from the SN rate as a function of redshift measured by Dahlen
et al. (2004) over the GOODS fields. Critical to this derivation is
the last data point at $z=1.6$. We show that the adopted extinction
effect at this redshift has been strongly underestimated compared to
actual measurements of the extinction for a representative sample of
galaxies at the same redshift derived from the GOODS database (Daddi
et al. 2007).  The assumption of extinction in high-redshift galaxies
being the same as locally clearly conflicts with all available
evidence, in particular when considering that galaxies at $z\gsim 1$
and $z\sim 2$ form stars at a factor of $\sim 10$ and $\sim 30$ higher
rate, respectively, compared to local galaxies (e.g. Elbaz et
al. 2007; Daddi et al. 2007) and have much higher gas fractions
(Bouch\'e et al. 2007; Daddi et al. 2008), and ULIRGs are $\sim 1000$
times more numerous than locally (Daddi et al. 2005). Thus, we
conclude that there is at present no physical ground for the Strolger
et al.  distribution of delay times, which suppress virtually all SNIa
events for delay times less than $\sim 2$ Gyr. Deep HST observations
in the near infrared are needed to properly measure the SNIa rate at
$z\gsim 1.4$.

At the opposite extreme, we examine the very {\it prompt}
distribution of delay times advocated by Mannucci et al. (2006), based
on a putative excess of 7 SNIa events in a sample of
radiogalaxies. This excess is attributed to minor episodes of star
formation concomitant with the activation of nuclear activity in
radiogalaxies.  We argue that if the 7 excess SNIa events were due to
recent bursts of star formation in radiogalaxies, then over 30 core
collapse supernovae should have exploded within the monitored sample of
radiogalaxies, but none was actually observed. We conclude that the
excess SNIa events are unlikely to be due to recent star formation,
and may just be the result of small number statistics, the excess
being indeed only a $2\sigma$ effect. Thus, there appears to be no
physical basis for the very prompt distribution of delay times
advocated by Mannucci et al. (2006).

Having excluded these mutually irreconcilable renditions of the SNIa
distribution of delay times, we recall that theoretically motivated
renditions for these distributions, for either the so-called single
degenerate (SD) or the so-called double degenerate (DD) channel, quite
naturally predict distributions of delay times with both a prompt as
well as a late component. Being these distribution controlled by
partly different physical effects in the SD and DD channels, the SD/DD
mix of SNIa's is predicted to vary in a systematic fashion as a
function of cosmic time (redshift). This effect is illustrated for
selected samples of SD and DD models, indicating that the SD/DD mix of
SNIa events is not expected to vary more than $\sim 50\%$ over an
extended period of cosmic times and redshifts. This is on the
reassuring side when using SNIa's as standard candles for the
determination of cosmological parameters.

\section*{Acknowledgments}

We thank an anonymous referee for constructive comments. AR acknowledges
the kind hospitality of the Osservatorio Astronomico di Padova.


\bigskip\noindent 
{\bf Note added in proof} It is possible that in Section 4 we have
overestimated the number of core collapse (CC) supernovae expected in
the sample of radiogalaxies if the 7 excess SNIa events are due to
recent star formation.  Della Valle et al. (2005) derive their
estimated number of CC events (0.62) originated from the postulated
star formation as the product of an expected rate of CC SNe (0.0152
$h^2_{75}$ SNu) times a control time (4057/100), and since star
formation was associated to radio loudness, we interpreted both
quantities as pertaining to the radio loud galaxies.  If instead rate
and control time were meant for the whole sample of ellipticals
(irrespective whether radio loud or not), the control time for radio
loud galaxies only should be lower than adopted in our calculation.
Della Valle et al. do not explicitly report the CC control time for
the radio loud galaxies, but to first order it is $(2199/11096)\times
4057=804$, where 2199 is the SNIa control time for radio loud
galaxies, and 11096 is that for the entire sample. This assumes the
fraction of control time for radio loud galaxies to be the same for
both CC and Type Ia supernove, which is justified by the observational 
database being the same for both SN types. If so, our
estimated number of expected CC events among radio loud galaxies (69)
has been overestimated by a factor $\sim 5$, hence should rather be
$\sim 14$ events, still $\sim 20$ times higher than estimated by Della
Valle et al (2005). We maintain our conclusion that the lack of observed
CC events remains a challenge for the interpretetation of the 7 excess SNIa 
events in terms of recent star formation. We are grateful to Massimo Della 
Valle for having pointed out to us our misinterpretation of their CC control 
time.  

\label{lastpage}

\end{document}